\documentclass[journal]{IEEEtran}
\IEEEoverridecommandlockouts
\usepackage{steinmetz}
\usepackage{hhline}
\usepackage{bm}
\usepackage{color}
\usepackage{longtable}	
\usepackage{multicol}
\usepackage{float}
\usepackage{multirow}
\usepackage{enumerate}
\usepackage{cite}
\usepackage{amsmath,amssymb,amsfonts}
\usepackage{algorithmic}
\usepackage{enumitem}
\usepackage{textcomp}
\usepackage{diagbox}
\usepackage{xcolor}
\usepackage{graphicx}
\usepackage{nicefrac}
\usepackage{physics}
\usepackage{array}
\newcolumntype{C}{>{\centering\arraybackslash}p{1.5cm}}

\ifCLASSOPTIONcompsoc
\usepackage[caption=false,font=normalsize,labelfon
t=sf,textfont=sf]{subfig}
\else
\usepackage[caption=false,font=footnotesize]{subfi
g}
\fi

\def\BibTeX{{\rm B\kern-.05em{\sc i\kern-.025em b}\kern-.08em
    T\kern-.1667em\lower.7ex\hbox{E}\kern-.125emX}}

  \usepackage{optidef}

    \usepackage{optidef}

\usepackage{algorithm}
\usepackage{algorithmic}
\definecolor{rred}{rgb}{0,0,0}
\usepackage{titlesec}
\renewcommand\thesubsubsection{\thesubsection.\arabic{subsection}}

\usepackage{mathtools}
\newtagform{rred}{\color{rred}(}{)}
\pdfoutput=1
\begin{document}

\IEEEoverridecommandlockouts

\title{An Online Network Model-Free Wide-Area Voltage Control Method Using PMUs\\
\thanks{The authors are with the Department of Electrical and Computer Engineering, McGill University, Montreal, QC H3A 0G4, Canada. (e-mail: georgia.pierrou@mail.mcgill.ca, xiaozhe.wang2@mcgill.ca)}
}

\author{Georgia~Pierrou,~\IEEEmembership{Student Member,~IEEE,} and~Xiaozhe~Wang,~\IEEEmembership{Senior Member,~IEEE}
\thanks{This work is supported by the Natural Sciences and Engineering Research Council (NSERC) under Discovery Grant NSERC RGPIN-2016-04570 and the Fonds de Recherche du Qu\'ebec-Nature et technologies under Grant FRQ-NT PR-253686.
}

}



\maketitle

\begin{abstract}
This paper proposes a novel online  measurement-based Wide-Area Voltage Control (WAVC) method using Phasor Measurement Unit (PMU) data in power systems with Flexible AC Transmission System (FACTS) devices. As opposed to previous WAVC methods, the proposed WAVC does not require any model knowledge or the participation of all buses and considers both active and reactive power perturbations. Specifically, the proposed WAVC method exploits the regression theorem of the Ornstein-Uhlenbeck process to estimate the sensitivity matrices through PMU data online, which are further used to design and apply the voltage regulation by updating the reference points of FACTS devices. Numerical results on the IEEE 39-Bus \textcolor{rred}{and IEEE 68-Bus systems} demonstrate that the proposed model-free WAVC can provide effective voltage control in various network topologies, different combinations of voltage-controlled and voltage-uncontrolled buses, under measurement noise, and in case of missing PMUs. Particularly, the proposed WAVC algorithm may outperform the model-based WAVC when an undetected topology change happens.
\end{abstract}

\vspace{-15pt}
\section{Introduction}

Voltage stability is crucial to ensure the normal operation of power systems. 
To avoid voltage instability/collapse, 
a classic voltage control scheme, also known as primary voltage control, is typically adopted, which utilizes the conventional generator Automatic Voltage Regulators (AVR) to maintain voltage levels at generator buses in case of local perturbations. 
However, the growing need to automatically coordinate the different control sources led to the secondary voltage control, which has been effectively implemented in French and Spanish power networks \cite{Paul87}, \cite{Sancha96}. Specifically, secondary voltage control aims to improve the overall voltage profile by controlling on-field reactive power resources, such as Flexible AC Transmission Systems (FACTS) devices, at some pilot buses \cite{Ilic95}.

In recent years, the increasing amount of Phasor Measurement Units (PMUs) provides a unique opportunity for the evolution of secondary voltage control. 
Wide-Area Voltage Controllers (WAVC) have been proposed by utilities (e.g., \cite{Taylor06, Perron17}) 
as a more sophisticated way to conduct the secondary voltage control using high-frequency, time-synchronized PMU  measurements. Indeed, different PMU-based WAVC methods can be found in the literature. In \cite{Liu10}, a PMU-based Automated Voltage Control and Automated Flow
Control is designed based on the optimal PMU location. An adaptive measurement-based voltage control algorithm that exploits measurements as feedback control inputs for FACTS devices is proposed in \cite{Su13}. 
The authors in \cite{Su18} present a nonlinear constrained optimization algorithm for voltage control that estimates the reactive power disturbances using PMUs online. In \cite{Musleh16}, a PMU-based WAVC algorithm is developed to control a FACTS device in real-time. An extension that takes into account the measurement time delay is proposed in \cite{Musleh18}.

Nonetheless, the WAVC formulations of 
all the aforementioned works strictly rely on network topology, as the susceptance matrix $B$ is needed. 
However, \color{rred}{grid-wide inter-area topology information is provided only on an hourly basis \cite{Zhu12}}. Thus, accurate network topology may not always be available. 
\textcolor{rred}{Indeed, topology errors have been recorded in the literature as a result of errors in the status of circuit breakers such as isolation switches \cite{Zhou18} or totally unobservable cyber attacks \cite{Zhang16}. }\color{black}  
Moreover, the aforementioned WAVC works applied the susceptance matrix $B$ to describe the sensitivity relation between the voltage deviation and the reactive power change, i.e., $\Delta \bm{Q}=J_{\bm{QV}}\Delta \bm{V} \approx B\Delta \bm{V}$, which neglects the impacts of real power perturbation on voltage magnitudes and is only true under the assumptions of flat voltages and the participation of \textit{all} buses in the voltage control.
However, such assumptions may not hold in practical applications due to various operating conditions, missing PMUs, and
limited communication links. 



In this paper, we 
propose a novel online model-free WAVC using PMU data, which is completely independent of model knowledge\textcolor{rred}{, topology information} and parameter values. Particularly, we regard the power system as a stochastic dynamic system and 
exploit the regression theorem of Ornstein–Uhlenbeck process to estimate the sensitivity matrices from PMU data, \textcolor{rred}{i.e.,  bus voltage magnitude and angle measurements,}  
in near real-time. Then the estimated sensitivity matrices are further used to
design and apply the WAVC 
by updating the reference points of SVCs, or other similar types of FACTS devices. 
 To the best of the authors' knowledge, this work seems to be the first attempt to develop a 
\textit{model-free} WAVC.
The main advantages of the proposed method are summarized as follows:
\begin{enumerate}
    \item The proposed WAVC algorithm, being model-free, can be implemented online to provide effective voltage control in various network topologies, operating conditions and different combinations of voltage-controlled and voltage-uncontrolled buses.
    \item The proposed WAVC method does not require the participation of all buses and thus provides more flexibility when practical constraints (e.g., missing PMUs, limited communication infrastructure) arise.
        \item  Unlike previous works, the proposed method provides effective voltage control considering both active and reactive power perturbations.
    \item Numerical studies show that the proposed WAVC remains effective under measurement noise and missing PMUs. The proposed method may even outperform the model-based WAVC if an undetected topology change occurs. 
\end{enumerate}

In addition, the model-free nature of the proposed method may offer additional flexibility in a decentralized framework in case of network partitioning into zones (e.g., \cite{Guo19, Yu12}), yet this paper will focus only on a centralized implementation.

\textcolor{rred}{Lastly, it is worth noting that, to achieve the estimation of the sensitivity matrices, a considerable amount of PMU data is required (300s measurements with a frequency of 60Hz in this paper), which nevertheless is still a reasonable time window for online application \cite{Shi20}. However, no network information and topology are needed. Thus, it is believed that the proposed model-free WAVC utilizing PMUs may compliment the traditional model-based WAVC to maintain the voltage level of power systems,  when accurate network information is not guaranteed.}

The remainder of the paper is structured as follows: Section \ref{1} introduces the stochastic dynamic power system model for voltage control. Section \ref{2} reviews the mathematical formulation for WAVC. In Section \ref{3}, the proposed model-free WAVC algorithm is elaborated. Section \ref{4} provides a detailed numerical study that demonstrates the effectiveness of the method. Section \ref{5} gives the conclusions. 

\vspace{-5pt}
\section{The Stochastic Dynamic Power System Model}
\label{1}

Voltage  stability   is  a dynamic phenomenon by nature.  
 Load dynamics and the associated control will greatly affect the voltage stability of a power system \cite{Cutsem}. Also, power systems are constantly experiencing small perturbations around the steady-state operating point, as a result of varying load and generation patterns. In view of this, in contrast to the static approach adopted by previous WAVC works, in this paper we consider 
 the inherent stochastic dynamics of power systems, which will be shown to provide an interesting way of extracting important information regarding the physical model from PMU measurements.
\vspace{-10pt}
\subsection{Load Dynamics}
\label{1a}
Load dynamics play a major role in voltage stability, while load buses equipped with Load Tap Changers (LTCs), FACTS etc. provide great flexibility in voltage control.
Therefore, load buses have been commonly selected as the main agents for voltage control (e.g., \cite{Ashrafi14}).
In this paper, we will design the WAVC at the load side based on the load dynamics and control. However, detailed generator dynamics are also incorporated in the simulation study in Section \ref{4} to demonstrate the efficacy of the proposed method in real-world applications.

Assuming that $m$ out of $L$ loads are dynamic loads that can be characterized by the dynamic load model proposed in \cite{Canizares:1995}:
\vspace{-10pt}
\begin{eqnarray}
\label{eq:dynload}
\dot{\theta}_{k} &=& \frac{1}{\tau_{\theta_k}}(P_k-P_k^s) \label{eq:conductance}\\
\dot{V}_k &=& \frac{1}{\tau_{V_k}}(Q_k-Q_k^s)\label{eq:susceptance}
\end{eqnarray}
\vspace{-20pt}
\begin{eqnarray}
\label{eq:loaddemand}
P_k = \sum_{j=1}^{N} V_kV_j(-G_{kj}\cos \theta_{kj}-B_{kj}\sin \theta_{kj})\\
Q_k = \sum_{j=1}^{N} V_kV_j(-G_{kj}\sin \theta_{kj}+B_{kj}\cos \theta_{kj})
\end{eqnarray}

\noindent where $k\in\{1,2,...,m\}$; $\theta_k$ $V_k$ are the bus voltage angle and magnitude, respectively; $\tau_{\theta_k}$, $\tau_{V_k}$ are the time constants of active and reactive power recovery; $P_k, Q_k$ are the active and reactive power absorptions in terms of the power injected from the network to the dynamic load buses; $P_k^s, Q_k^s$ are the steady-state active and reactive power absorptions.



The applied load model can adequately characterize a variety of load types (e.g., thermostatic loads, induction motors, loads controlled by LTCs) in \textcolor{rred}{the} voltage stability study, whose difference lies in different 
values of the time constants $\tau_{\theta_k}$ and $\tau_{V_k}$ that may range from  milliseconds to several minutes. The static loads can also be naturally represented by taking the limit $\tau_{\theta_k}\rightarrow 0$, $\tau_{V_k}\rightarrow 0$ \cite{Nguyen16}. Similar types of load are also introduced in \cite{Liu92, Overbye94,Nguyen16}. The applied model captures the qualitative load behavior over a wide range of voltage magnitudes observed in the voltage stability study, where voltage magnitudes will respond to load power 
to maintain reactive power balance \cite{DeMarco90}. \color{black}

Random load variations may be caused by the aggregate behavior of individual users \cite{Nguyen16}. To model load fluctuations, as a common and reasonable approach \cite{Nwankpa93}, we add Gaussian stochastic perturbations to the steady-state power: 
\vspace{-3pt}
\begin{eqnarray}
\label{eq:stochdynload1}
d\theta_k &=& \frac{1}{\tau_{\theta_k}}(P_k-P_k^s)dt-\frac{1}{\tau_{\theta_k}}P_k^s\sigma_{k}^Pd\xi_k^{P} \\
dV_k &=& \frac{1}{\tau_{V_k}}(Q_k-Q_k^s)dt-\frac{1}{\tau_{V_k}}Q_k^s\sigma_{k}^Qd\xi_k^{Q}
\label{eq:stochdynload2}
\end{eqnarray}
where $\sigma_{k}^P, \sigma_{k}^Q$ describe the standard deviation of the stochastic load perturbation for the active and reactive power; $\xi_k^P, \xi_k^Q$ are Wiener processes.

Equations (\ref{eq:stochdynload1})-(\ref{eq:stochdynload2}) can be linearized and written in the following vector form:
\vspace{-4pt}
\usetagform{rred}
\begin{eqnarray}
\begin{bmatrix} {d\bm{\theta}} \\ d\bm{V}
\end{bmatrix} &=&
\begin{bmatrix}T_{\theta}^{-1}&\bm{0}\\\bm{0}&T_{V}^{-1}\end{bmatrix}
\begin{bmatrix} \frac{\partial \bm{P}}{\partial \bm{\theta}} & \frac{\partial \bm{P}}{\partial \bm{V}}\\ \frac{\partial \bm{Q}}{\partial \bm{\theta}}& \frac{\partial \bm{Q}}{\partial \bm{V}}  \end{bmatrix}
\begin{bmatrix} \bm{\theta} \\ \bm{V} \end{bmatrix}dt \nonumber \\ &+& \begin{bmatrix} -T_{\theta}^{-1}P^s\Sigma^P  & \bm{0} \\ \bm{0} &  -T_{V}^{-1}Q^s\Sigma^Q  \end{bmatrix} \begin{bmatrix}
d\bm{\xi^P} \\ d\bm{\xi^Q}  \end{bmatrix}
\label{eq:matrixload}
\end{eqnarray}
 \usetagform{default}
where
\begin{tabular}{ l l }
 & \\
  $\bm{\theta} = \begin{bmatrix} \theta_{1}, ..., \theta_{m} \end{bmatrix}^T$,  & $\bm{V} = \begin{bmatrix} V_{1}, ..., V_{m} \end{bmatrix}^T$, \\
  $T_{\theta} = $ diag $\begin{bmatrix} \tau_{\theta_1}, ..., \tau_{\theta_m} \end{bmatrix}$, & $T_{V} = $ diag $\begin{bmatrix} \tau_{V_1}, ..., \tau_{V_m} \end{bmatrix}$, \\
  $\bm{P}= \begin{bmatrix} P_{1}, ..., P_{m} \end{bmatrix}^T$, &
  $\bm{Q}= \begin{bmatrix} Q_{1}, ..., Q_{m} \end{bmatrix}^T$,\\
  $P^s= $ diag $\begin{bmatrix} P_{1}^s, ..., P_{m}^s \end{bmatrix}$, &  $Q^s= $ diag $ \begin{bmatrix} Q_{1}^s, ..., Q_{m}^s \end{bmatrix}$, \\
  $\Sigma^P = $ diag $\begin{bmatrix} \sigma_{1}^P, ..., \sigma_{m}^P \end{bmatrix}$, & $\Sigma^Q = $ diag $\begin{bmatrix} \sigma_{1}^Q, ..., \sigma_{m}^Q \end{bmatrix}$,\\
  $\bm{\xi^P} = \begin{bmatrix} \xi_{1}^P, ..., \xi_{m}^P \end{bmatrix}^T$, &
  $\bm{\xi^Q} = \begin{bmatrix} \xi_{1}^Q, ..., \xi_{m}^Q \end{bmatrix}^T$
\end{tabular}\\

We denote $\bm{x}= \begin{bmatrix} \bm{\theta}, \bm{V} \end{bmatrix}^T $, $A = \begin{bmatrix} T_{\theta}^{-1}\frac{\partial \bm{P}}{\partial \bm{\theta}} & T_{\theta}^{-1}\frac{\partial \bm{P}}{\partial \bm{V}}\\ T_{V}^{-1}\frac{\partial \bm{Q}}{\partial \bm{\theta}}&  T_{V}^{-1}\frac{\partial \bm{Q}}{\partial \bm{V}}  \end{bmatrix}$, $ H=\begin{bmatrix} -T_{\theta}^{-1}P^s\Sigma^P  & \bm{0} \\ \bm{0} &  -T_{V}^{-1}Q^s\Sigma^Q  \end{bmatrix}$, $\bm{\xi}=\begin{bmatrix} \bm{\xi^P}, \bm{\xi^Q}  \end{bmatrix}^T$, then (\ref{eq:matrixload}) takes the following compact form:
\begin{equation}
    \label{eq:OUprocess}
    d\bm{x}=A\bm{x}dt+H d\bm{\xi}
\end{equation}
Hence, the stochastic dynamic load model in ambient conditions can be represented as a vector Ornstein-Uhlenbeck process \cite{Gardiner}. The system state matrix $A$ corresponds to a scaled set of sensitivity matrices and carries significant information of the system operating state, as will be shown in Section \ref{2}. In Section \ref{3}, we will propose a purely data-driven method to estimate the matrix $A$, which is further utilized to develop the WAVC algorithm.  

\vspace{-10pt}
\subsection{SVC Modeling}
FACTS devices have been widely used for wide-area control with various control objectives. Among the most commonly used FACTS devices for reactive power compensation and voltage control is the Static VAR Compensator (SVC), which is a thyristor controlled reactor-based shunt FACTS device \cite{Gurrola14}. \color{black} In this paper, SVCs are assumed to be installed at some dynamic load buses to perform the voltage control. In other words, the dynamic load buses with SVCs are termed as voltage-controlled buses. The following dynamic model is used for the SVC \cite{Acha}: 
\begin{eqnarray}
\label{eq:VM_svc}
\dot{V}_{M} &=& \frac{1}{T_M}(K_MV_{k}-V_M)\\
\label{eq:alpha_svc}
\dot{\alpha} &=& \frac{1}{T_2}(-K_D\alpha+K\frac{T_1}{T_2T_M}(V_M-K_MV_k)) \nonumber \\&+& \frac{K}{T_2}(V_{k,ref}-V_M)
\end{eqnarray}
where $\alpha$ is the firing angle; $V_{M}$ is the filtered voltage at bus $k$; $V_{k}$ is the voltage magnitude at bus $k$;  $V_{k,ref}$ is the reference voltage at bus $k$; $K_M, K_D, K$ are the regulator gains; $T_M, T_1, T_2$ are the regulator time constants. 

The reactive power injected at the voltage-controlled bus $k$ where the SVC is connected can be described as:
\begin{equation}
\label{eq:svcpower}
Q_{SVC}=\frac{2\alpha-\sin2\alpha-\pi(2-x_L/x_C)}{\pi x_L}V_{k}^2
\end{equation}
where $x_L$ is the SVC inductive reactance and $x_C$ is the SVC capacitive reactance.

An important advantage of the SVC controller is that, except for the reactive power support, it can be tuned to directly control the voltage of load bus $k$ by adjusting the set-point $V_{k,ref}$. It is worth noting that similar voltage control methods may be developed using different types of FACTS devices \textcolor{rred}{that operate based on a reference voltage setting,} such as Static Synchronous Compensators (STATCOM), as implemented in \cite{Musleh18}, \cite{Nguyen05}. \color{black}
\vspace{-5pt}
\section{Wide-Area Voltage Control}\label{2}
\subsection{Mathematical Formulation}
\label{2a}

When the power system is in normal operating conditions, the linearized power flow model can be applied.  
At the dynamic load buses, we have:
\begin{equation}
\begin{bmatrix} \Delta \bm{P} \\ \Delta \bm{Q}
\end{bmatrix} = \begin{bmatrix}  J_{\bm{P}\bm{\theta}} &  J_{\bm{P}\bm{V}}\\ J_{\bm{Q}\bm{\theta}}&  J_{\bm{Q}\bm{V}}\end{bmatrix} \begin{bmatrix}
 \Delta \bm{\theta}  \\ \Delta\bm{V} \end{bmatrix} =
J
\begin{bmatrix} \Delta \bm{\theta}  \\ \Delta\bm{V} \end{bmatrix}
\label{eq:powerflow}
\end{equation}
where $\Delta\bm{P}=[\Delta P_{1}, ..., \Delta P_{m}]^T$ is the vector of active power changes; $\Delta\bm{Q}=[\Delta Q_{1}, ..., \Delta Q_{m}]^T$ is the vector of reactive power changes; $\Delta\bm{\theta}=[\Delta \theta_{1}, ..., \Delta \theta_{m}]^T$ is the vector of bus voltage angle changes; $\Delta\bm{V}=[\Delta V_{1}, ..., \Delta V_{m}]^T$ is the vector of bus voltage magnitude changes at the dynamic load buses; $J_{\bm{P}\bm{\theta}}=\pdv{\bm{P}}{\bm{\theta}}, J_{\bm{P}\bm{V}}=\pdv{\bm{P}}{\bm{V}}, J_{\bm{Q}\bm{\theta}}=\pdv{\bm{Q}}{\bm{\theta}}, J_{\bm{Q}\bm{V}}=\pdv{\bm{Q}}{\bm{V}}$.
Equation (\ref{eq:powerflow}) can also be written as follows:

\begin{equation}
\begin{bmatrix} \Delta \bm{\theta} \\ \Delta \bm{V}
\end{bmatrix} =
\begin{bmatrix}  S_{\bm{\theta}\bm{P}} &  S_{\bm{\theta}\bm{Q}}\\ S_{\bm{V}\bm{P}}&  S_{\bm{V}\bm{Q}}\end{bmatrix}
\begin{bmatrix} \Delta \bm{P}  \\ \Delta\bm{Q} \end{bmatrix} = S
\begin{bmatrix} \Delta \bm{P}  \\ \Delta\bm{Q} \end{bmatrix}
\label{eq:powerflow_inv}
\end{equation}
where $S=J^{-1}$. 
Therefore, we can obtain the following voltage control model:
\begin{eqnarray}
\Delta\bm{V}=S_{\bm{V}\bm{P}} \Delta \bm{P}+S_{\bm{V}\bm{Q}} \Delta \bm{Q}
\label{eq:matrixB1}
\end{eqnarray}
In previous WAVC literature\cite{Liu10, Su13, Su18, Musleh16, Musleh18}, the decoupled power flow \cite{Grainger} formulation  $\Delta \bm{Q}=J_{\bm{Q}\bm{V}} \Delta \bm{V}$ or $\Delta \bm{V}=S_{\bm{V}\bm{Q}} \Delta \bm{Q}$  is applied by neglecting the impacts of real power perturbation on voltage deviation. In contrast, this paper considers the impacts of  both active and reactive power mismatches on voltage deviations.

\color{black}
\vspace{-10pt}
\subsection{Wide-Area Voltage Control}
\label{2b}
In this work, we consider the scenario that SVCs are available at some dynamic load buses that could perform the voltage control so as to minimize the voltage deviation at the other load buses that are without the capability of voltage regulation.
Therefore, (\ref{eq:matrixB1}) can be written as:
\begin{eqnarray}
\begin{bmatrix} \Delta\bm{V_{c}} \\ \Delta\bm{V_{u}}
\end{bmatrix} &=&  \nonumber
\begin{bmatrix}  S_{\bm{VP}_{cc}} & S_{\bm{VP}_{cu}}\\ S_{\bm{VP}_{uc}}& S_{\bm{VP}_{uu}}\end{bmatrix}
\begin{bmatrix} \Delta\bm{P_{c}}  \\ \Delta\bm{P_{u}} \end{bmatrix}\\
 &+&\begin{bmatrix}  S_{\bm{VQ}_{cc}} & S_{\bm{VQ}_{cu}}\\ S_{\bm{VQ}_{uc}}& S_{\bm{VQ}_{uu}}\end{bmatrix}
\begin{bmatrix} \Delta\bm{Q_{c}}  \\ \Delta\bm{Q_{u}}\end{bmatrix}
\label{eq:matrixB}
\end{eqnarray}
where the subscripts $c$ and $u$ denote the $n_{c}$ voltage-controlled load
buses with SVCs installed and the $n_u$ voltage-uncontrolled load
buses, respectively,  
while $n_c+n_u=m$. Note that the submatrices $S_{\bm{VP}_{cc}}, S_{\bm{VP}_{cu}}, S_{\bm{VP}_{uc}}, S_{\bm{VP}_{uu}}$, $S_{\bm{VQ}_{cc}}, S_{\bm{VQ}_{cu}}, S_{\bm{VQ}_{uc}}, S_{\bm{VQ}_{uu}}$ are obtained by reordering the rows and columns of the matrices $S_{\bm{VP}}, S_{\bm{VQ}}$ according to the selection of the controlled and uncontrolled load buses. The voltage deviation is defined as 
$\Delta\bm{V}=\bm{V}-\bm{V_{ref}}$, where $\Delta\bm{V}=[
\Delta\bm{V_c}, \Delta\bm{V_u}]^T$, $\bm{V}=[
\bm{V_c}, \bm{V_u}
]^T$ and $\bm{V_{ref}}=[
\bm{V_{c, ref}}, \bm{V_{u, ref}}]^T$. Particularly, for voltage-controlled buses, $\bm{V_{c, ref}}$ denote the SVC reference voltages that can be adjusted. 
For voltage-uncontrolled buses, $\bm{V_{u, ref}}$ is constant and corresponds to 
the steady-state power flow solution. Also, since SVCs are used at the voltage-controlled buses, $\Delta\bm{P_{c}}$ is assumed to be $\bm{0}$. 
After some algebraic manipulation of (\ref{eq:matrixB}), the following relations can also be derived:
\begin{eqnarray}
 \Delta\bm{Q_{c}} = S_{\bm{VQ}_{cc}}^{-1} [  \Delta\bm{V_{c}} -S_{\bm{VP}_{cu}} \Delta\bm{P_{u}}-S_{\bm{VQ}_{cu}} \Delta\bm{Q_{u}}]
 \label{eq:deltaQc}
\\
 \Delta\bm{V_{u}}= S_{\bm{VP}_{uu}} \Delta\bm{P_{u}} + S_{\bm{VQ}_{uc}} \Delta\bm{Q_{c}}+S_{\bm{VQ}_{uu}} \Delta\bm{Q_{u}}
\label{eq:deltaQu_deltaVu}
\end{eqnarray}

By substituting (\ref{eq:deltaQc}) into (\ref{eq:deltaQu_deltaVu}), the voltage deviation of the uncontrolled buses can be written as:
\begin{eqnarray}
 \Delta\bm{V_{u}}
 &=&  \nonumber [S_{\bm{VP}_{uu}}-S_{\bm{VQ}_{uc}}S_{\bm{VQ}_{cc}}^{-1}S_{\bm{VP}_{cu}}]\Delta\bm{P_{u}} \\  &+& \nonumber [S_{\bm{VQ}_{uu}}-S_{\bm{VQ}_{uc}}S_{\bm{VQ}_{cc}}^{-1}S_{\bm{VQ}_{cu}}]\Delta\bm{Q_{u}}\\&+&
S_{\bm{VQ}_{uc}}S_{\bm{VQ}_{cc}}^{-1}\Delta\bm{V_{c}}
\label{eq:deltaVu}
\end{eqnarray}
Assuming that some active and reactive power perturbations $\Delta\bm{P_{u}}(t_i)$ and $\Delta\bm{Q_{u}}(t_i)$ occur at uncontrolled buses at the current time step $t_i$, the goal of the voltage controller is to minimize the voltage deviation of the uncontrolled buses $\Delta\bm{V_{u}}(t_{i+1})$ at the next time step $t_{i+1}=t_i+ \Delta t$, by adjusting the reference points of the SVCs at the controlled buses through $\Delta\bm{V_{c}}(t_{i+1})$.   

Hence,  the online optimization problem can be formulated as:  
\begin{equation}
 \label{eq:minimizationProblem}
\hspace{-0.2pt}\begin{array}{rrclcl}
\displaystyle \min_{\Delta \bm{V_c}(t_{i+1})} & \multicolumn{1}{l}{||\Delta\bm{V_u}(t_{i+1})||_{\infty}}\\
\textrm{s.t.} & \bm{V_{c}^{min}}(t_{i+1}) \leq \bm{V_{c}}(t_{i+1}) \leq \bm{V_{c}^{max}}(t_{i+1}) \\
&\bm{Q_{c}^{min}}(t_{i+1}) \leq \bm{Q_{c}}(t_{i+1}) \leq \bm{Q_{c}^{max}}(t_{i+1})  \\
\end{array}
\end{equation}



In previous WAVC works \cite{Liu10, Su13, Su18, Musleh16, Musleh18}, the impacts of real power perturbation on voltage deviation are not considered, i.e., $\Delta \bm{Q}=J_{\bm{QV}} \Delta \bm{V}$ or $\Delta \bm{V}=S_{\bm{V}\bm{Q}} \Delta \bm{Q}$, have been used. Besides, the approximation  $J_{\bm{Q}\bm{V}}\approx B$ is applied in which $B$ is assumed to be fully known in real-time from network topology and line parameters.
However, as mentioned in the Introduction, accurate topology information may not always be available due to telemetry failure, bad data, undetected topology change, etc. 
In addition, 
the assumption $J_{\bm{Q}\bm{V}}\approx B$ is true only under the condition that the voltage magnitudes of all buses are available 
i.e., $\Delta \bm{Q}$ and $\Delta \bm{V}$ in the approximation $\Delta \bm{Q}=B\Delta \bm{V}$ 
have to include all buses.
Nevertheless, such assumption may be rarely met in real life due to missing PMUs, limited communication links, etc. 

In this paper, we will propose a model-free WAVC method that is independent of network topology and can provide effective voltage control considering both active and reactive power perturbations ($\Delta \bm{P_{u}}, \Delta \bm{Q_{u}}$ in (\ref{eq:deltaVu})) in various working conditions, network topologies, and in case of missing PMUs.


\vspace{-5pt}
\section{A Model-Free Wide-Area Voltage Control Method}
\label{3}
\subsection{A Model-Free Method to Estimate the Sensitivity Matrices} \label{3A}
\label{3a}
In Section \ref{1a}, we have shown that the stochastic dynamic load modeling corresponds to a vector Ornstein-Uhlenbeck process (see (\ref{eq:OUprocess})). The stationary covariance matrix $C_{\bm{xx}}$ of the process is defined as follows:
\vspace{-2pt}
\begin{equation}
\label{eq:covariance}
C_{\bm{xx}} = \big \langle [\bm{x}(t)-\mu_{\bm{x}}][\bm{x}(t)-\mu_{\bm{x}}]^T \big  \rangle = \begin{bmatrix}
C_{\bm{\theta\theta}} & C_{\bm{\theta V}} \\
C_{\bm{V\theta}} &   C_{\bm{VV}}
\end{bmatrix}
\end{equation}
where $\mu_{\bm{x}}$ is the mean of the process. The $\tau$-lag autocorrelation matrix is defined as:
\vspace{-2pt}
\begin{equation}
\label{eq:tlag_corelation}
G(\tau) = \big \langle [\bm{x}(t+\tau)-\mu_{\bm{x}}][\bm{x}(t)-\mu_{\bm{x}}]^T \big  \rangle
\end{equation}
Since the normal operation of a power system around its steady state is considered, the system state matrix $A$ is stable.
By the \textit{regression theorem} of the vector Ornstein-Uhlenbeck process \textcolor{rred}{which holds under ambient conditions} \cite{Gardiner}, the $\tau$-lag autocorrelation matrix satisfies the following matrix differential equation:
\begin{equation}
\label{eq:regressiontheorem}
\frac{d}{d\tau} [G(\tau)] = AG(\tau)
\end{equation}
where $G(0)=C_{\bm{xx}}$. As a result, the dynamic state matrix $A$ can be obtained by solving (\ref{eq:regressiontheorem}) \cite{Sheng19}:
\begin{equation}
\label{eq:matrixA_regressiontheorem}
A=\frac{1}{\tau}\log\begin{bmatrix}G(\tau)C_{\bm{xx}}^{-1}\end{bmatrix}
\end{equation}
Equation (\ref{eq:matrixA_regressiontheorem}) tactfully relates the statistical properties of PMU measurements 
with the physical properties of the dynamic power system model and provides an intriguing way of estimating the system state matrix from PMUs. The regression theorem of the vector Ornstein-Uhlenbeck process has been leveraged in \cite{Sheng19} to estimate the system state matrix of classical model of generators, which can be used in mode identification, stability monitoring, etc. However, the relation (\ref{eq:matrixA_regressiontheorem}) deduced from the regression theorem is, for the first time, applied to extract the sensitivity matrices  for voltage control, which will be shown to be accurate and effective.

In practical applications, $G(\tau)$ and $C_{\bm{xx}}$ can be estimated from PMU measurements. Assuming that a window size of $n$ PMU measurements $\bm{x}(t_i)=[\bm{\theta}(t_i), \bm{V}(t_i)]^T, i=1,2,...,n$ are available at the dynamic load buses, 
the sample mean $\bm{\bar{x}}$, the sample covariance matrix $\hat{C}$ and the sample $\tau$-lag correlation matrix $\hat{G}$ can be calculated as follows:
\begin{equation}
\label{eq:xbar}
\bar{\bm{x}}= \frac{1}{n}\sum_{i=1}^{n} \bm{x}(t_i)
\end{equation}
\begin{equation}
\label{eq:Chat}
\hat{C}= \frac{1}{n-1}(F-\bar{\bm{x}}\bm{1}_{n}^T)(F-\bar{\bm{x}}\bm{1}_{n}^T)^T
\end{equation}
\vspace{-4pt}
\begin{equation}
\label{eq:tlaghat}
\hat{G}(\Delta t)= \frac{1}{n-1}(F_{2:n}-\bar{\bm{x}}\bm{1}_{n-1}^T)(F_{1:n-1}-\bar{\bm{x}}\bm{1}_{n-1}^T)^T
\end{equation}
where
$F=\begin{bmatrix} \bm{x}(t_1), ...,\bm{x}(t_n) \end{bmatrix}$ is a $2m \times n$ matrix with the states' measurements, $\bm{1}_{n}$ is a $n \times 1$ vector of ones, $\Delta t$ is the time lag, $F_{i:j}$ denotes the submatrix of $F$ from the $i$-th to the $j$-th column. Hence, the estimated dynamic system state matrix $\hat{A}$ can be obtained as:
\begin{equation}
    \label{eq:Ahat}
    \hat{A}=\frac{1}{\Delta t}\log \begin{bmatrix} \hat{G}(\Delta t)\hat{C}^{-1}\end{bmatrix}=\begin{bmatrix} \hat{T}_{\theta}^{-1}\hat{J}_{\bm{P}\bm{\theta}}& \hat{T}_{\theta}^{-1}\hat{J}_{\bm{P}\bm{V}}\\ \hat{T}_{V}^{-1}\hat{J}_{\bm{Q}\bm{\theta}}&  \hat{T}_{V}^{-1}\hat{J}_{\bm{Q}\bm{V}} \end{bmatrix}
\end{equation}
where $A$ is a scaled Jacobian matrix (see (\ref{eq:OUprocess})).
Particularly, the submatrices of $\hat{A}$ correspond to an estimation of the scaled sensitivity matrices used in the WAVC method discussed in Section \ref{2}. If the time constants $T_\theta, T_V$  can be estimated, the sensitivity matrices 
needed for the WAVC can be extracted purely from PMU measurements without any knowledge of the network model, which is required by previous WAVC works. 
\vspace{-10pt}
\subsection{Estimating the Load Time Constants from PMUs}
\label{3b}
By looking at the dynamic load  model (\ref{eq:conductance})-(\ref{eq:susceptance}), one can easily observe that when PMU measurements for  $\bm{\theta}, \bm{V}$ and $\bm{P}, \bm{Q}$ are available, the time constants $\tau_{\theta_k}, \tau_{V_k}$  at each dynamic load bus $k=1,2,...,m$ can be estimated through a simple linear regression analysis \cite{Miller}, because we have:
\begin{eqnarray}
  \frac{\Delta{\theta}_k}{\Delta t} &=& \frac{1}{\hat{\tau}_{\theta_k}}(P_k-P_k^s)\\
    \frac{\Delta{V}_k}{\Delta t} &=& \frac{1}{\hat{\tau}_{V_k}}(Q_k-Q_k^s)
\end{eqnarray}
For this purpose, we denote the angle and voltage deviations at each load bus $k$ between the consecutive samples $i$, $i-1$ as  $\Delta {\theta}_k(t_i)= {\theta}_k(t_i)-{\theta}_k(t_{i-1}), \Delta {V}_k(t_i)= {V}_k(t_i)-{V}_k(t_{i-1})$, where $i=1,2,...,p$. Besides, we denote the deviation of active and reactive power absorption as $\Delta {P}_k(t_i)={P}_k(t_i) - \bar{{P}}_k, \Delta {Q}_k(t_i)={Q}_k(t_i) - \bar{{Q}}_k$, where $\bar{{P}}_k, \bar{{Q}}_k$ are the sample means, i.e. $\bar{{P}}_k= \frac{1}{p}\sum_{i=1}^{p} {P}_k(t_i), \bar{{Q}}_k= \frac{1}{p}\sum_{i=1}^{p} {Q}_k(t_i)$.
Then the estimator for each load time constant should minimize the residuals \begin{eqnarray}
  \hat{\varepsilon}_k^{\tau_{\theta_k}}(t_i)&=& \frac{\Delta \theta_k(t_i)}{\Delta t}-\frac{1}{\hat{\tau}_{\theta_k}} \Delta P_k(t_i)  \\
     \label{eq:T_k_error}
     \hat{\varepsilon}_k^{\tau_{V_k}}(t_i)&=& \frac{\Delta V_k(t_i)}{\Delta t}-\frac{1}{\hat{\tau}_{V_k}} \Delta Q_k(t_i)
\end{eqnarray}
So $\hat{T}_{\theta} = $ diag $\begin{bmatrix} \hat{\tau}_{\theta_1}, ..., \hat{\tau}_{\theta_m} \end{bmatrix}, \hat{T}_{V} = $ diag $\begin{bmatrix} \hat{\tau}_{V_1}, ..., \hat{\tau}_{V_m} \end{bmatrix}$ can be obtained by solving
the following minimization problems:
\begin{equation}
 \label{eq:T_estimation_2}
\begin{array}{rrclcl}
\displaystyle \min_{\hat{\tau}_{\theta_k}} & \multicolumn{3}{l}{\sum_{i=1}^p \hat{\varepsilon}_k^{\tau_{\theta_k}}(t_i)^2}\\
\end{array}
\end{equation}
\begin{equation}
 \label{eq:T_estimation}
\begin{array}{rrclcl}
\displaystyle \min_{\hat{\tau}_{V_k}} & \multicolumn{3}{l}{\sum_{i=1}^p \hat{\varepsilon}_k^{\tau_{V_k}}(t_i)^2}\\
\end{array}
\end{equation}

Once the time constants  $\hat{T}_\theta, \hat{T}_V$ are estimated, the sensitivity matrices $\hat{J}_{\bm{P}\bm{\theta}}, \hat{J}_{\bm{P}\bm{V}}, \hat{J}_{\bm{Q}\bm{\theta}}, \hat{J}_{\bm{Q}\bm{V}}$ can be extracted from the scaled ones estimated in Section \ref{3a} and further be exploited in the design of the model-free WAVC.
\vspace{-10pt}
\subsection{The Proposed Model-Free WAVC Algorithm}
\label{3c}
\begin{figure}[!htb]
\centering
\includegraphics[width=2.2in ,keepaspectratio=true,angle=0]{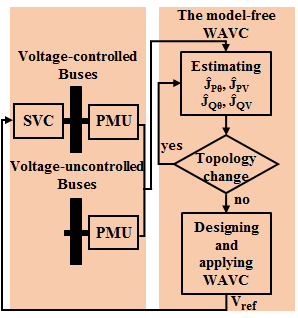}
\caption{An illustration of the proposed model-free WAVC.}
\label{wavc_fig}
\vspace{-15pt}
\end{figure}

We have seen from the previous discussions that the sensitivity
matrices $\hat{J}_{\bm{P}\bm{\theta}}, \hat{J}_{\bm{P}\bm{V}}, \hat{J}_{\bm{Q}\bm{\theta}}, \hat{J}_{\bm{Q}\bm{V}}$ and the time constants $\hat{T}_\theta, \hat{T}_V$ can be estimated purely from PMU data (Section \ref{3a} and \ref{3b}) and can be further exploited in the design of WAVC (Section \ref{2}). As such, a novel online model-free WAVC method can be developed to minimize the voltage deviation of the uncontrolled load buses without any model information. The algorithm is summarized below.
Particularly, \textbf{Step E1}-\textbf{Step E4} are for the estimation of  the sensitivity matrices, while \textbf{Step C1}-\textbf{Step C4} are for the online WAVC. The structure of the proposed model-free WAVC is also illustrated in Fig. \ref{wavc_fig}. 

\begin{figure}[!htb]
\centering
\includegraphics[width=3.6in ,keepaspectratio=true,angle=0]{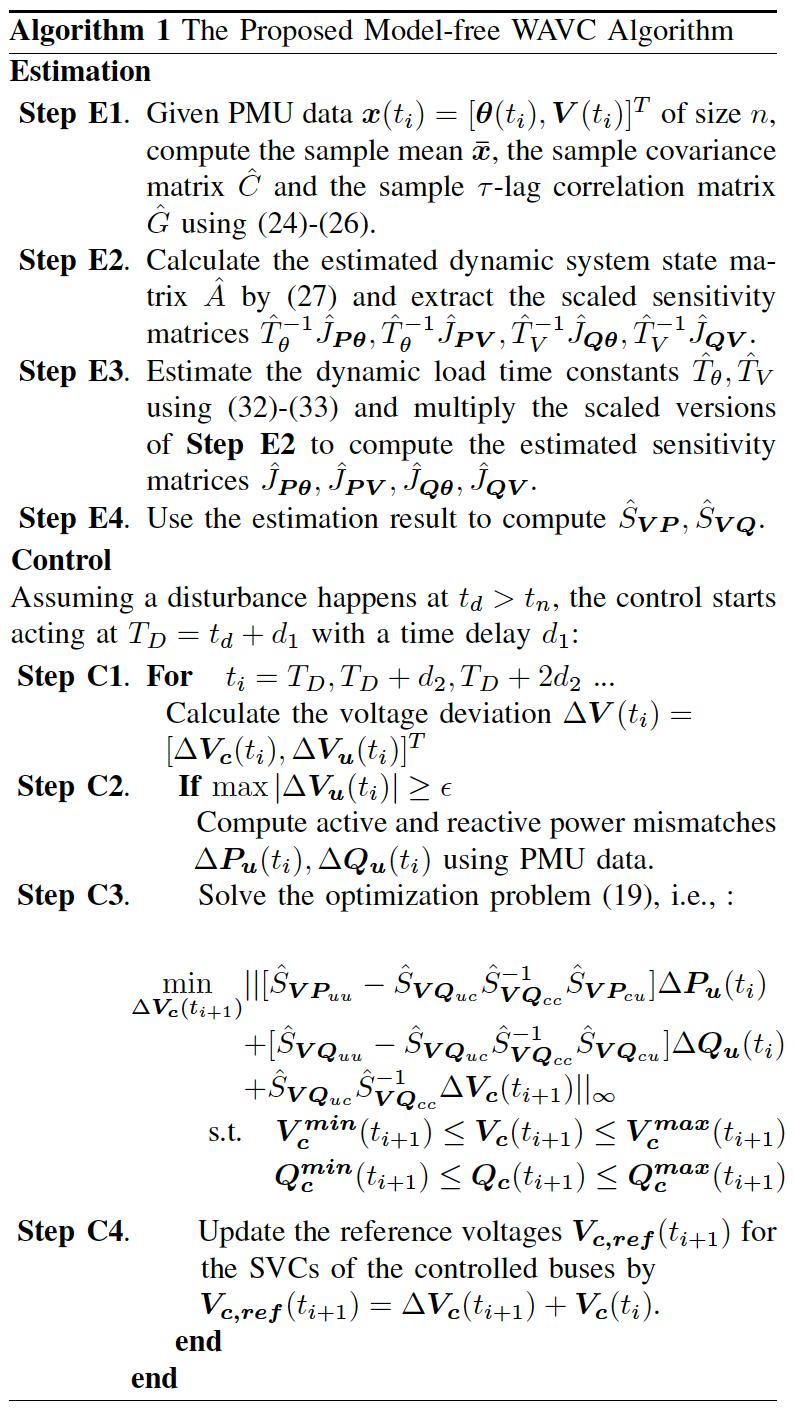}
\vspace{-15pt}
\end{figure}

 Remarks:\\
 \noindent $\bullet$  In this paper, a window size of 300s with a sampling frequency of 60 Hz is used in \textbf{Step E1}, for which a good accuracy is achieved. It is worth noting that, despite of a relatively large window size,  the above algorithm does not assume any network model information (e.g. topology, line parameters), but can estimate the sensitivity matrices purely from PMU data. 

\noindent $\bullet$ In case the algorithm detects changes in network topology that may significantly affect the sensitivity matrices, new PMU data needs to be collected when the system settles to the new steady state to reestimate the sensitivity matrices and redesign the WAVC, i.e., \textbf{Step E1-Step E4} should be repeated. \textcolor{rred}{Alternatively, to achieve more accurate and frequent estimation and control, the current algorithm using one window of measurements can be extended to an online recursive algorithm using a moving window, similar to the approach in \cite{Pierrou20}.}

\noindent $\bullet$ Similar to the approach adopted in \cite{Ashrafi14}, the proposed WAVC is triggered with a delay $d_1$ after a disturbance to avoid excessive interaction between the control and the transient response of voltages. $d_1=30s$ in the simulation studies of this paper, whereas different values may be selected based on different systems' dynamic characteristics.

\noindent  $\bullet$ The time interval between two control actions $d_2$ is selected to be 0.2s (e.g., 10 sampling time steps), similar to the value chosen in  \cite{Musleh18}, to quickly mitigate the voltage deviation due to disturbance. 
However, a larger time interval can also be selected to 
help avoid conflict with other control mechanisms, as described in \cite{Ashrafi14}. 


\noindent  $\bullet$ The proposed method assumes that PMU measurements are available at all dynamic load buses.  However, as will be shown in Section \ref{4b}, the model-free WAVC remains effective in case of \color{rred}some \color{black} missing measurements at either voltage-controlled or voltage-uncontrolled dynamic load buses. \textcolor{rred}{It is worth mentioning that it is beyond the scope of the paper to to deal with the occurrence of 
severe bad data events. Bad data detection methods \cite{Kekatos13}, \cite{Filho14} and missing data recovery methods \cite{Osipov20, Hao18} can be applied to pre-process the PMU data before applying the proposed model-free WAVC algorithm. } 




\vspace{-5pt}
\section{Numerical Results}
\label{4}

The IEEE 39-Bus \textcolor{rred}{and IEEE 68-Bus systems have} been used to validate and test the proposed model-free WAVC. 
The nonlinear programming solver \textit{fmincon} and the interior-point algorithm \cite{Musleh18} are used to get the optimal solution at \textbf{Step C3} of the proposed algorithm. The simulation study was implemented in PSAT-2.1.10 \cite{Milano05}.

To quantify the performance of the proposed model-free WAVC, 
the root mean square value of the voltage deviations at the $n_u$ uncontrolled load buses in steady state can be used as the performance index $\lambda$ \cite{Su18}, i.e.,:
\begin{equation}
     \lambda = \sqrt{\frac{1}{n_u} ||\Delta{\bm{V_u}}^{ss}||^2_{2}}\\
\end{equation}
where 
$\Delta\bm{V_u}^{ss}=\bm{V_u}^{ss}-\bm{V_{u, ref}}$. $\bm{V_u}^{ss}$ are the new steady-state voltages of the voltage-uncontrolled buses; $\bm{V_{u, ref}}$ are the previous steady-state voltages of those buses before the perturbation occurs.
\vspace{-10pt}
\textcolor{rred}{\subsection{Validation on the IEEE 39-Bus Test System}\label{4a}}
In this section, the validation of the proposed algorithm is conducted on the IEEE 39-Bus test system. Both load dynamics and generator dynamics are considered in the numerical study. Particularly, the 4th-order generator models equipped with AVRs are included and 19 loads are modelled as the stochastic dynamic loads described in (\ref{eq:stochdynload1})-(\ref{eq:stochdynload2}). 
The time constants  $T_\theta, T_V$ are set to be 
30s. The load model considers Gaussian stochastic load variations with a mean of zero and $\sigma_{k}^P=\sigma_{k}^Q=1$ in (\ref{eq:stochdynload1})-(\ref{eq:stochdynload2}). SVCs are installed at 3 voltage-controlled buses (Bus 3, 9, 20), while the rest 16 load buses are voltage-uncontrolled. Different combinations of controlled and uncontrolled buses will be discussed later.

\begin{figure}[!tb]
\centering
\includegraphics[width=3.5in ,keepaspectratio=true,angle=0]{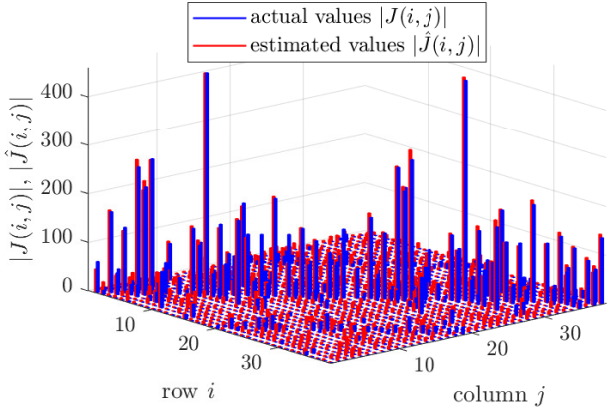}
\caption{The estimation results for matrix $J$ of the IEEE 39-Bus system.}
\label{esterror_ieee39}
\end{figure}

Firstly, \textcolor{rred}{the \textbf{Estimation} procedure in \textbf{Algorithm 1} is followed and} the accuracy of the estimated sensitivity matrix is demonstrated. To this end, 300s PMU data are collected from the 19 load buses, from which the sample covariance matrix, the sample $\tau$-lag correlation matrix, and the scaled sensitivity matrices $\hat{T}_\theta^{-1}\hat{J}_{\bm{P}\bm{\theta}}, \hat{T}_\theta^{-1}\hat{J}_{\bm{P}\bm{V}}, \hat{T}_V^{-1}\hat{J}_{\bm{Q}\bm{\theta}}, \hat{T}_V^{-1}\hat{J}_{\bm{Q}\bm{V}}$ are estimated (\textbf{Step E1-Step E2}). Moving to \textbf{Step E3-Step E4}, all time constants  in $\hat{T}_\theta, \hat{T}_V$ are well estimated from 0.1s (among the collected 300s) PMU data by the linear regression analysis, with acceptable estimation errors less than $10\%$. Next, the sensitivity matrices $\hat{J}_{\bm{P \theta}}, \hat{J}_{\bm{PV}}, \hat{J}_{\bm{Q \theta}}, \hat{J}_{\bm{QV}}$ and $\hat{S}_{\bm{VP}}, \hat{S}_{\bm{VQ}}$ can be obtained. Specifically, Fig. \ref{esterror_ieee39} presents a comparison between the estimated and the true values of matrix $J$, evidently showing the good accuracy of the estimation. 

\textcolor{rred}{Once the estimated sensitivity matrices are obtained, the WAVC presented in the \textbf{Control} part of \textbf{Algorithm 1} can be designed and activated in case of some perturbation. To test the effectiveness of the proposed model-free WAVC, we regard that the estimation result is obtained at $t_n=0$s and a 25$\%$ load increase is applied to the active and reactive power of all voltage-uncontrolled buses at $t_d=2$s}. 
The WAVC starts working with a delay of $d_1=30$s to avoid excessive interaction of the controller during the transient. The controller is executed every $d_2=0.2s$ if $\max |\Delta\bm{V_{u}}(t_i)| \geq 0.005 \mbox{p.u.}$.

\begin{figure}[!tb]
\centering\includegraphics[width=2.3in ,keepaspectratio=true,angle=0]{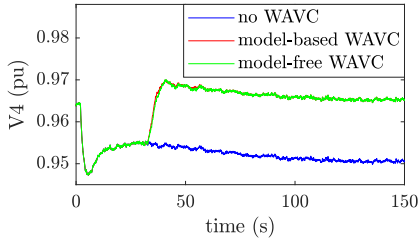}
\vspace{-5pt}
\caption{The voltage profile at Bus 4 with 3 voltage-controlled buses (Case C - Bus 3, 9, 20).}
\label{validation1_fig}
 \vspace{-5pt}
\end{figure}

 Fig. \ref{validation1_fig} depicts a comparison of the voltage magnitude at the uncontrolled Bus 4 for different control scenarios (no control, model-based WAVC using the true matrix $J$, the proposed model-free WAVC using the estimated matrix $\hat{J}$). As can be seen, the model-free WAVC is as effective as the model-based WAVC, as both can boost up the voltage when 
 load power increases. In addition, the performance index $\lambda$ for the different scenarios of no control, model-based WAVC using the true matrix $J$ and the proposed model-free WAVC assuming 3 voltage-controlled buses can be found in the third row of Table \ref{table:validation} (Case C). These results clearly show the good accuracy of the estimated sensitivity matrices and the effectiveness of the proposed model-free WAVC. The rest of the voltage-uncontrolled dynamic load buses exhibit similar behaviors. 

In terms of efficiency, using a Windows computer with a quad-core Intel i7 2.2 GHz processor and 8GB RAM, \textbf{Step E1-Step E4} took 0.312775s for estimating the sensitivity matrices, while the solution of the optimization problem \textbf{Step C3} took much less than the time interval between two control actions (e.g. only 0.059321s). The fast computational speed clearly demonstrates the good feasibility of the proposed WAVC in online implementation.

In the above simulations, 3 voltage-controlled buses and 16 voltage-uncontrolled buses are assumed. Nevertheless, different combinations of voltage-controlled and voltage-uncontrolled buses may affect the performance of the proposed model-free WAVC. To investigate this, 
we test the performance of the proposed method in different combinations of controlled and uncontrolled buses. Specifically, 4 different cases are tested where the number of voltage-controlled buses varies from 1 (Case A - Bus 3) to 4 (Case D - Bus 3, 9, 20, 23), as 
shown in Table \ref{table:validation}. Note that for each case, a  25$\%$ increase in the active and  reactive power consumption is applied at the corresponding voltage-uncontrolled buses that are different from Case A to Case D, i.e., the load increasing pattern varies in different cases. It is evident from Table \ref{table:validation} that, similar to the performance of the control using the true sensitivity matrices, the proposed model-free WAVC can also provide effective control in all cases.




\color{blue}
\begin{table}[!t]
\centering
  \caption{The performance index $\lambda$ for various combinations of controlled and uncontrolled buses}\label{table:validation}
   \setlength{\tabcolsep}{2pt}
  \begin{tabular}{|c|c|c|c|c|}
\hhline{|-|-|-|-|-|}
\parbox[c]{.1\linewidth}{\centering \vspace{10pt} \textbf{Case}}&\parbox[c]{0.3\linewidth}{ \vspace{10pt}\centering \textbf{Voltage-controlled Buses}}  &
\multicolumn{3}{c|}{\begin{tabular}{c} $\bm{\lambda}$ \end{tabular}} \\ \cline{3-5}
 &  & \begin{tabular}{c} \textbf{no WAVC} \end{tabular} & \begin{tabular}{c} \textbf{model-based}\\ \textbf{WAVC} \end{tabular} & \begin{tabular}{c}\textbf{model-free} \\ \textbf{WAVC} \end{tabular} \\ \hline
A&3&0.012737&0.0080188&0.0080193
\\
\hline
B&3, 20&0.010373&0.0048502&0.0048502\\
\hline
C&3, 9, 20& 0.01113&0.0040667&0.0040674\\
\hline
D&3, 9, 20, 23&0.010072&0.0040306&0.0040308\\

\hhline{|-|-|-|-|-|}
  \end{tabular}
  \vspace{-10pt}
\end{table}
\color{black}

\subsection{\color{rred}Impact of Missing PMUs and PMU locations}
\label{4b}
\color{black}
Although the proposed model-free WAVC assumes that PMU measurements are available at all dynamic load buses as stated in Sections \ref{2}-\ref{3}, this assumption may not always hold 
in practical applications. In this section, we validate the performance of the proposed methodology in case of missing PMUs at either voltage-controlled or voltage-uncontrolled buses. It should be noted that due to the loss of PMU measurements, Equation (\ref{eq:matrixB}) should be modified as follows:
\begin{eqnarray}
\begin{bmatrix} \Delta\bm{V_{c}}' \\ \Delta\bm{V_{u}}'
\end{bmatrix} &=&  \nonumber
\begin{bmatrix}  S_{\bm{VP}_{cc}}' & S_{\bm{VP}_{cu}}'\\ S_{\bm{VP}_{uc}}'& S_{\bm{VP}_{uu}}'\end{bmatrix}
\begin{bmatrix} \Delta\bm{P_{c}}'  \\ \Delta\bm{P_{u}}' \end{bmatrix}\\
 &+&\begin{bmatrix}  S_{\bm{VQ}_{cc}}' & S_{\bm{VQ}_{cu}}'\\ S_{\bm{VQ}_{uc}}'& S_{\bm{VQ}_{uu}}'\end{bmatrix}
\begin{bmatrix} \Delta\bm{Q_{c}}'  \\ \Delta\bm{Q_{u}}'\end{bmatrix}
\label{eq:matrixB'}
\end{eqnarray}
where $'$ is used to highlight that buses with missing PMUs are excluded from the mathematical formulation of the controller. Specifically,  $S_{\bm{VQ}_{cc}}', S_{\bm{VQ}_{cu}}', S_{\bm{VQ}_{uc}}', S_{\bm{VQ}_{uu}}'$
are obtained from $S_{\bm{VQ}_{cc}}, S_{\bm{VQ}_{cu}}, S_{\bm{VQ}_{uc}}, S_{\bm{VQ}_{uu}}$ by omitting the rows and columns corresponding to the buses with missing PMUs. A similar operation applies to $S_{\bm{VP}}$. \textcolor{rred}{Thus, in case of PMU loss, the \textbf{Estimation} procedure in \textbf{Algorithm 1} can still estimate the sensitivity matrices at the dynamic load buses with PMUs available. Regarding the \textbf{Control} procedure in \textbf{Algorithm 1},} when PMU measurements are missed at previously voltage-controlled buses, those buses no longer receive the WAVC signals or updates to their SVCs' reference points. Similarly, the voltage deviation of voltage-uncontrolled buses not equipped with PMUs is no longer considered while solving \textbf{Step C3}.


\textcolor{rred}{To study the impact of missing measurements, we consider some of the PMUs of Case C to be lost.  \color{rred}To determine the minimum number of PMUs available, as discussed in \cite{Ree10}, PMUs should be placed in about one third of the buses to achieve full observability of the system. Therefore, we assume that PMUs are installed at (1/3)*39=13 dynamic load buses in the IEEE 39 bus test system. Previously, we had 19 PMUs, now we have 13 PMUs available, i.e., 6 missing PMUs.}

\begin{figure}[!t]
\centering
\vspace{-5pt}
\includegraphics[width=2.3in ,keepaspectratio=true,angle=0]{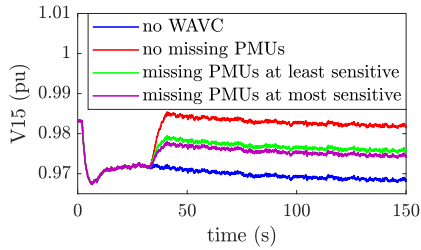}
\caption{\color{rred}The voltage profile at Bus 15 in case of missing PMUs with 3 voltage-controlled buses (Case C - Bus 3, 9, 20).\color{black}}
\vspace{-5pt}
\label{missingpmu_fig_rev}
\end{figure}

\color{rred}In addition to missing PMUs, the installation locations of available PMUs may also affect the performance of estimation and control. Many previous works have studied the optimal locations of PMUs \cite{Nuqui05, Chakrabarti08, Makram11}. Particularly, we follow 
a sensitivity analysis approach as in \cite{Makram11}. 
A sensitivity indicator is defined as the power flow rise $\Delta |S_{ij}|$ at each transmission line $i-j$ following a 25$\%$ load power increase $\Delta \bm{P}$ at the dynamic load buses, i.e., $\frac{\Delta |S_{ij}|}{\Delta \bm{P}}$. It is recommended in \cite{Makram11} that available PMUs should be located at the dynamic load buses connecting the most sensitive transmission lines, i.e., the ones that experience the largest power flow rises. In our case, assuming we have only 13 PMUs available, we compare the performance of the proposed method under two cases of missing PMUs:
\begin{itemize}
    \item Case C.I (best case): The 13 available PMUs are placed at the most sensitive dynamic load buses connecting the top 13 sensitive transmission lines (Bus 3, 4, 7, 8, 12, 15, 16, 18, 23, 24, 25, 26, 28), whereas 6 missing PMUs are considered at the least sensitives buses (Bus 1, 9, 20, 21, 27, 29).
        \item Case C.II (worst case):The 13 available PMUs are placed at the least sensitive buses (Bus 1, 8, 9, 12, 16, 20, 21, 23, 24, 26, 27, 28, 29) and missing PMUs are considered at the most sensitive buses (Bus 3, 4, 7, 15, 18, 25).
\end{itemize}
\color{black}\color{rred}Once the sensitivity matrices at the dynamic load buses with PMU data available are estimated, the same load disturbance as in Section \ref{4a} applies. Fig. \ref{missingpmu_fig_rev} presents the voltage profile of the uncontrolled Bus 15 for the different cases. It can be seen that a large number of missing PMUs may deteriorate the performance of voltage control compared to the case with no missing PMUs. However, the sensitivity analysis may contribute to the optimal placement of PMUs and thus improve the performance of the voltage control. With missing PMUs at the least sensitive buses (Case C.I, the best case), the performance index is only slightly affected comparing to the case with no missing PMUs, but is much better than that in the case where missing PMUs happen at the most sensitive buses (Case C.II, worst case), as demonstrated in Table \ref{table:missingPMU_rev}.
\color{black}
The above results confirm the robustness of the proposed algorithm against missing measurements, which further reinforces its overall performance and feasibility in practical applications.






\begin{table}[!t]
\centering
\caption{\color{rred}The impact of missing PMUs and PMU locations on the performance index $\lambda$}
\setlength{\tabcolsep}{1.5pt}
\begin{tabular}{|c | c | c|c|c | c |}
\hhline{|-|-|-|-|-|-|}
 \parbox[c]{.1\linewidth}{\centering \vspace{12pt} \textbf{Case}}&\parbox[c]{0.2\linewidth}{\vspace{12pt} \centering \textbf{Voltage-controlled Buses}}  &
\multicolumn{4}{c|}{\begin{tabular}{c} $\bm{\lambda}$  \end{tabular}} \\ \cline{3-6}
 &  & \begin{tabular}{c}\color{rred} \textbf{no WAVC} \end{tabular} &\color{rred} \begin{tabular}{c}  \textbf{no missing}\\ \textbf{PMUs} \end{tabular} & \color{rred} \begin{tabular}{c}  \textbf{missing}\\ \textbf{PMUs} \\ \textbf{at least}\\ \textbf{sensitive} \end{tabular} & \color{rred} \begin{tabular}{c}  \textbf{missing}\\ \textbf{PMUs}  \\ \textbf{at most}\\ \textbf{sensitive} \end{tabular}  \\ \hline
C&3, 9, 20&0.01113&0.0040674& \color{rred}0.0058466\color{black}& \color{rred}0.0074339\color{black}\\
\hhline{|-|-|-|-|-|-|}
\end{tabular}
\vspace{-10pt}
\label{table:missingPMU_rev}

\end{table}

\vspace{-10pt}
\subsection{Impact of Measurement Noise}
The accuracy of measurement-based methods may be deteriorated by 
measurement noise. In this section, we test the performance of the proposed model-free WAVC against measurement noise.  \textcolor{rred}{For this purpose, measurement noise is added to the gathered PMU data of $\bm{\theta}, \bm{V}$ in \textbf{Step E1} of the \textbf{Estimation} part as well as to the PMU data of $\bm{V}$ in \textbf{Step C1} of the \textbf{Control} part in \textbf{Algorithm 1}. Two different types of noise, representing low and high noise levels, are considered.} Particularly, for the low noise level, we follow the approach in \cite{Zhou13} to add Gaussian noise with zero mean and standard deviation equal to 10\% of the largest state changes to the PMU measurements of $\bm{\theta}$ and $\bm{V}$. On the other hand, high noise levels correspond to Gaussian noise with fixed standard deviation $10^{-4}$ to the measurements of $\bm{\theta}$ and $\bm{V}$ \cite{Brown16}. \textcolor{rred}{Once the sensitivity matrices are estimated from the PMU data with added noise, a 25$\%$ increase in the active and reactive power of the voltage-uncontrolled buses is applied for Case C of Section \ref{4a}}. Table \ref{table:mn} depicts the values of the performance index $\lambda$ for the two noise levels. It can be observed that both noise levels lead to only insignificant changes to the performance index $\lambda$, demonstrating the robustness of the proposed method against measurement noise.


\begin{table}[!t]

\caption{The impact of measurement noise on the performance index $\lambda$}
\centering
\setlength{\tabcolsep}{1.5pt}
\begin{tabular}{|c | c | c|c|c |}
\hhline{|-|-|-|-|-|}
 \parbox[c]{.1\linewidth}{\centering \vspace{10pt} \textbf{Case}}&\parbox[c]{0.3\linewidth}{ \vspace{10pt}\centering \textbf{Voltage-controlled Buses}}  &
\multicolumn{3}{c|}{\begin{tabular}{c} $\bm{\lambda}$\\ \textbf{model-free} \textbf{WAVC} \end{tabular}} \\ \cline{3-5}
 &  & \begin{tabular}{c} \textbf{no noise} \end{tabular} & \begin{tabular}{c} \textbf{low noise} \end{tabular} & \begin{tabular}{c}\textbf{high noise} \end{tabular}  \\ \hline

C&3, 9, 20&0.0040674& 0.0040673&0.0045575\\
\hhline{|-|-|-|-|-|}
\end{tabular}
\label{table:mn}
\vspace{-10pt}
\end{table}


\vspace{-10pt}
\subsection{Effectiveness Under Topology Change}

One salient feature of the proposed online model-free WAVC method is its independency of network model such that it can adaptively update the estimation of the sensitivity matrices $J_{\bm{P \theta}}, J_{\bm{PV}}, J_{\bm{Q \theta}}, J_{\bm{QV}}$ and the control signals as the power grid evolves. Particularly, if an undetected topology change occurs in the system, the proposed method may outperform the model-based WAVC, the performance of which greatly relies on an accurate network model. To show this,
we consider that in Case E with 5 voltage-controlled buses, a three-phase fault occurs at Bus 26 that lasts for 10 cycles. To clear the fault, the breaker between Bus 26 and Bus 27 trips, which is unfortunately undetected by the topology processor. 

\textcolor{rred}  {
After the topology change occurs, a transient period follows and the system eventually settles down to a new steady-state where the application of the regression theorem for the estimation of the sensitivity matrix is allowed.}
\textcolor{rred}{Once the system reaches a new steady state after the topology change, new PMU data of 300s are used to estimate the new sensitivity matrices in $\hat{J} \prime$ in the \textbf{Estimation} part of \textbf{Algorithm 1}. The updated estimated sensitivity matrices are assumed to be available at $t_n=0$s and they are further used to design and apply the WAVC in the \textbf{Control} part of \textbf{Algorithm 1}.
To validate the effectiveness of the model-free WAVC after the topology change, a $30\%$ increase in the reactive power absorption of the voltage-uncontrolled buses is applied in the new steady state at $t_d=20$s, whereas the controller is activated with a delay of $d_1=30$s, i.e. at $T_D=50$s.}

\begin{figure}[!t]
\centering
\includegraphics[width=2.3in ,keepaspectratio=true,angle=0]{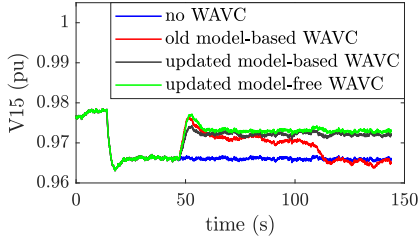}
\vspace{-5pt}
\color{rred}\caption{The voltage profile at Bus 15 in case of a line outage with 5 voltage-controlled buses (Case E - Bus 3, 9, 12, 20, 23).}\color{black}
\vspace{-5pt}
\label{fault_fig}
\end{figure}

In Fig. \ref{fault_fig}, the voltage profile at the voltage-uncontrolled Bus 15 is presented. Thanks to the new estimated sensitivity matrices, the proposed model-free WAVC is still able to 
effectively minimize the voltage deviation despite the topology change. However, the model-based WAVC is unable to perform the voltage control properly, as the sensitivity matrices are not updated after the undetected line outage. 
The advantage of the proposed model-free WAVC is also confirmed in Table \ref{table:gridcont}, where the values of the performance index $\lambda$ in different cases are presented. It can be observed that the proposed online model-free WAVC can effectively reduce the voltage deviations in different topologies \textcolor{rred}{and its performance is very similar to the updated model-based method, highlighting the accuracy of the estimation after a topology change}.

\begin{table}[!t]
\color{rred}\caption{The performance index $\lambda$ when using different sensitivity matrices under topology change}
\color{black}
\centering
\setlength{\tabcolsep}{1.5pt}
\begin{tabular}{|c | c | c|c|c | c |}
\hhline{|-|-|-|-|-|-|}
 \parbox[c]{.07\linewidth}{\centering \vspace{12pt} \textbf{Case}}&\parbox[c]{0.2\linewidth}{\vspace{12pt} \centering \textbf{Voltage-controlled Buses}}  &
\multicolumn{4}{c|}{\begin{tabular}{c} $\bm{\lambda}$  \end{tabular}} \\ \cline{3-6}
 &  & \begin{tabular}{c}\color{rred} \textbf{no WAVC} \end{tabular} &\color{rred} \begin{tabular}{c} \textbf{old}\\ \textbf{model-based}\\ \textbf{WAVC} \end{tabular} & \color{rred}\begin{tabular}{c} \textbf{updated}\\ \textbf{model-based}\\ \textbf{WAVC} \color{black}  \end{tabular} & \begin{tabular}{c} \textbf{updated}\\\textbf{model-free}\\\textbf{WAVC} \end{tabular}  \\ \hline
E&3, 9, 12, 20, 23&0.0089529&0.012143&\color{rred}0.0045857&\color{black}0.0047066\\
\hhline{|-|-|-|-|-|-|}
\end{tabular}
\vspace{-10pt}
\label{table:gridcont}
\end{table}
\vspace{-5pt}

\color{rred}
\vspace{-5pt}
\subsection{Validation on the IEEE 68-Bus System}
To show the feasibility of the proposed method in practical large-scale power systems, 
a numerical study on the IEEE 68-Bus test system is conducted. In this case, 
35 dynamic loads are considered with time constants of 30s. Gaussian stochastic load variations with a mean of zero and $\sigma_{k}^P=\sigma_{k}^Q=1$ are applied in (\ref{eq:stochdynload1})-(\ref{eq:stochdynload2}). SVCs are installed at 5 voltage-controlled buses (Bus 20, 25, 29, 41, 42).

\begin{figure}[!hbt]
\centering
\includegraphics[width=3.5in ,keepaspectratio=true,angle=0]{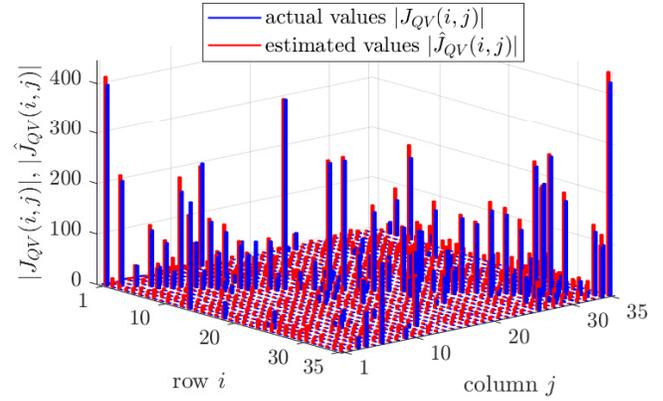}
\vspace{-10pt}
\caption{\textcolor{rred}{The estimation results for the sensitivity matrix $J_{\bm{QV}}$ of the IEEE 68-Bus system.}}
\label{esterror_ieee68}
\end{figure}

To apply the proposed WAVC method, 300s PMU measurements are collected to estimate the sensitivity matrices in $J$. 
A comparison between the elements of the true and the estimated sensitivity matrix $J_{\bm{QV}}$ is presented in Fig. \ref{esterror_ieee68}, showing a good accuracy of the estimation. 

\begin{figure}[!hbt]
\centering
\vspace{-10pt}
\includegraphics[width=2.3in ,keepaspectratio=true,angle=0]{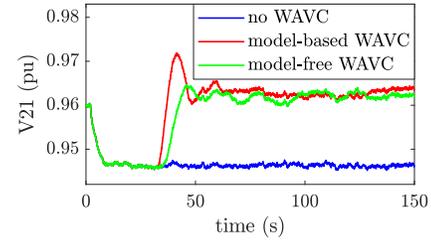}
\vspace{-5pt}
\caption{\textcolor{rred}{The voltage profile at Bus 21 with 5 voltage-controlled buses (Bus 20, 25, 29, 41, 42).}}
\vspace{-10pt}
\label{validation_68bus_fig1}
\end{figure}

To test the performance of the proposed WAVC, 
a 20$\%$ increase of the reactive power at voltage-uncontrolled buses is applied at $t_d = 2$s. The controller is activated with a delay $d_1=30s$ and once  $\max |\Delta\bm{V_{u}}(t_i)| \geq 0.01$p.u. at some uncontrolled bus. Fig. \ref{validation_68bus_fig1} shows the time evolution of the voltage magnitude at the voltage-uncontrolled Bus 21. It can be observed that, as with the control using the true sensitivity matrix, the proposed algorithm can effectively restore the voltages  
using the estimated matrix $\hat{J}_{\bm{QV}}$ in the larger test system. 

\vspace{-10pt}
\color{rred}
\subsection{Comparison to Other Methods}

\color{rred}
Although the proposed method seems to be the first purely measurement-based WAVC by exploiting the estimated sensitivity matrices, there there are previous works on estimating sensitivity matrices from PMUs \cite{Chen16, Li18}.
In this section, we compare the performance of the proposed estimation method and that of the Least Squares (LS) and Total Least Squares (TLS) methods developed in \cite{Chen16} in estimating the sensitivity matrices. Specifically, we compare the results of the proposed method and the TLS method under measurement noise.

\begin{figure}[!b]
\centering
\includegraphics[width=3.5in ,keepaspectratio=true,angle=0]{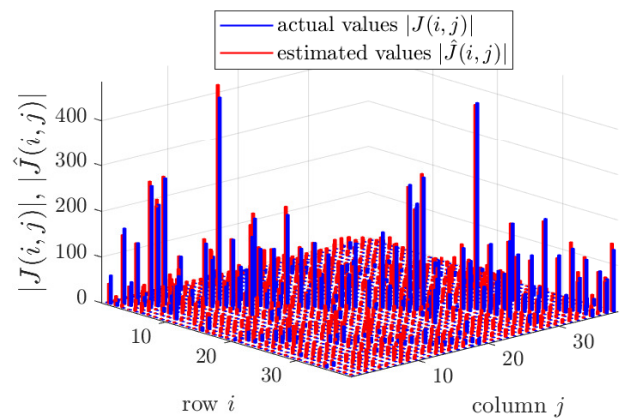}
\caption{\color{rred}The estimation result for matrix $J$ of the IEEE 39-Bus system using the proposed method under high measurement noise levels.\color{black}}
\label{ourmethod_paper}
\end{figure}

\begin{figure}[!tb]
\centering
\includegraphics[width=3.5in ,keepaspectratio=true,angle=0]{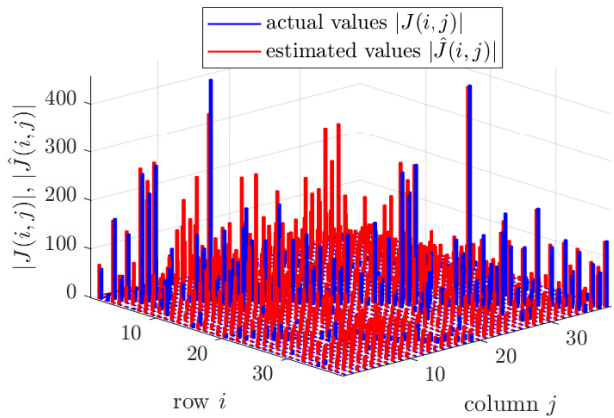}
\caption{\color{rred}The estimation result for matrix $J$ of the IEEE 39-Bus system using the TLS method \cite{Chen16} under high measurement noise levels.\color{black}}
\label{tlsmethod_paper}
\end{figure}

As have been shown in Section V-C (also validated in Fig. \ref{ourmethod_paper}), the impact of high measurement noise levels (i.e., Gaussian measurement noise with standard deviation $10^{-4}$ to the measurements of $\bm{\theta},\bm{V}$) on the performance of the proposed method is very small. On the other hand, Fig. \ref{tlsmethod_paper} shows that measurement noise can significantly
 deteriorate the estimation result using the TLS method. Similar results are observed when using the LS method. It is to note that other observations are also acquired from the comparison: 1). The estimation error using the LS and TLS methods strongly depends on the selected samples, as different samples may give very different results; 2). The TLS method may suffer from overfitting problem \cite{Davila93}. Thus, it may be difficult to determine the number of samples needed to achieve a satisfactory estimation result, which in turn may hamper its online implementation. 

\color{black}
\vspace{-13pt}
\section{Conclusions}
\vspace{-3pt}
\label{5}
Leveraging on the inherent power system dynamics, this paper has proposed a \textit{model-free} WAVC {for the first time}. 
Specifically, the proposed method can minimize voltage deviation by controlling the available FACTS devices in an online environment, without knowing the topology and line parameters of power network. 
The proposed WAVC method also provides more flexibility regarding the number and combination of buses participating in the voltage control scheme and unlike previous works, considers both active and reactive power perturbations.
Numerical results on the IEEE 39-Bus \color{rred}and 68-Bus \color{black}systems show that the proposed online model-free WAVC can provide effective voltage control in various network topologies, different combinations of voltage-controlled and voltage-uncontrolled buses, under measurement noise, and in case of missing PMUs. In addition, the proposed model-free WAVC may outperform the model-based WAVC if an undetected topology change occurs.



\begin{IEEEbiography}[{\includegraphics[width=1in,height=1.25in,clip,keepaspectratio]{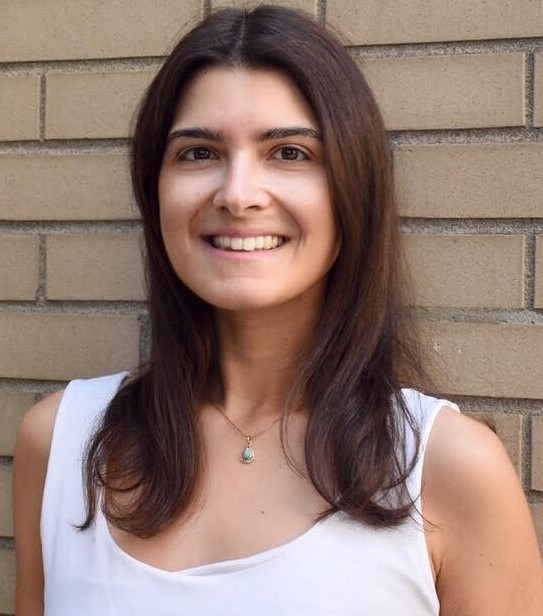}}]%
{Georgia Pierrou}
(S'19) received a Diploma in Electrical and Computer Engineering from the National Technical University of Athens, Athens, Greece in 2017. Since 2017, she has been
pursuing the Ph.D. degree in Electrical Engineering with the Electric Energy Systems Laboratory at McGill University, Montreal, Canada. Her research interests include power system dynamics, control and uncertainty quantification.
\end{IEEEbiography}

\begin{IEEEbiography}[{\includegraphics[width=1in,height=1.25in,clip,keepaspectratio]{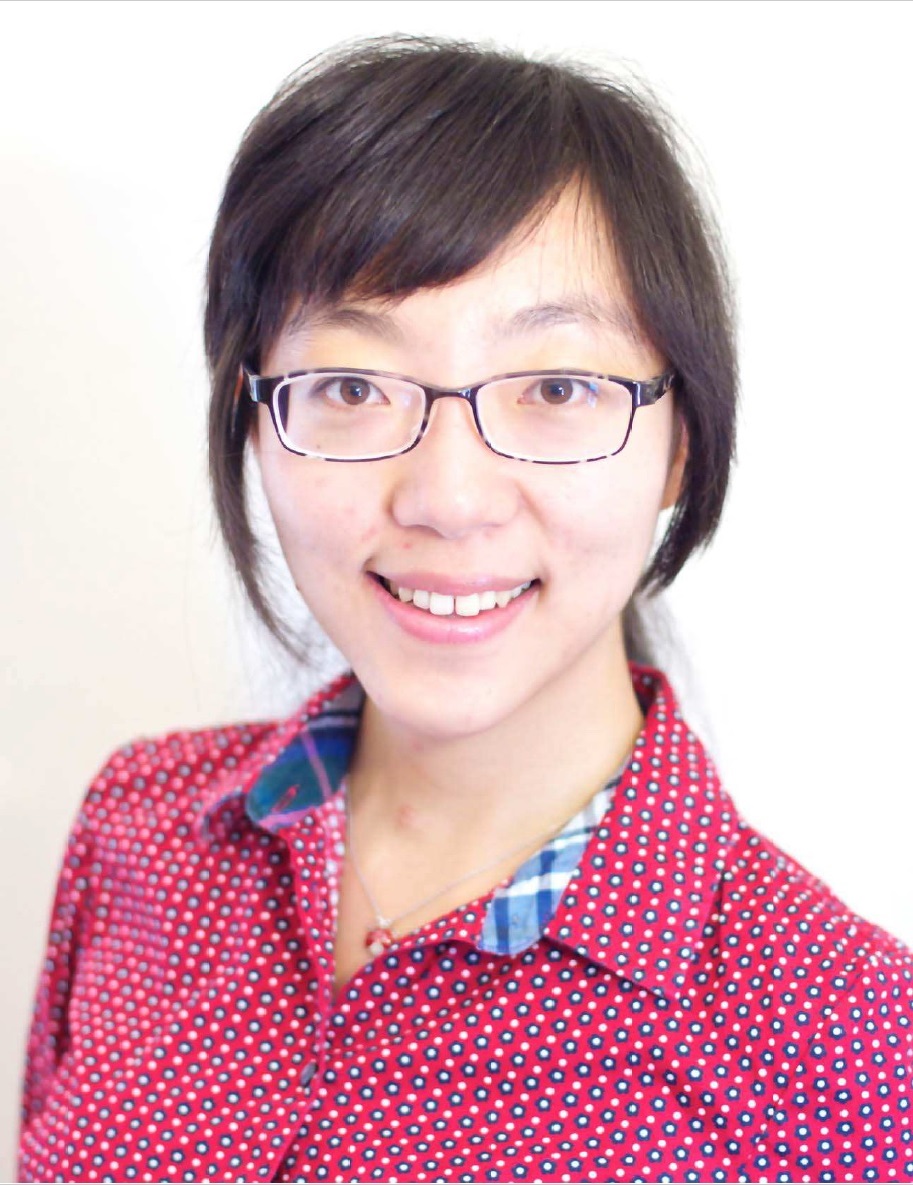}}]%
{Xiaozhe Wang}
(S'13-M'15-SM'20) is currently an Assistant Professor in the Department of Electrical and Computer Engineering at McGill University, Montreal, QC, Canada. She received the Ph.D. degree in the School of Electrical and Computer Engineering from Cornell University, Ithaca, NY, USA, in 2015, and the B.S. degree in Information Science \& Electronic Engineering from Zhejiang University, Zhejiang, China, in 2010. Her research interests are in the general areas of power system stability and control, uncertainty quantification in power system security and stability, and wide-area measurement system (WAMS)-based detection, estimation, and control. She is an IEEE Senior Member, serving on the editorial boards of IEEE Transactions on Power Systems, Power Engineering Letters, and IET Generation, Transmission and Distribution.
\end{IEEEbiography}

\end{document}